\documentclass[12pt]{iopart}
\usepackage{epsf}
\usepackage{iopams}
\usepackage{graphicx, wrapfig}
\usepackage{array}
\usepackage{amssymb}
\usepackage{setstack}
\usepackage{bm}
\usepackage{color}
\usepackage{mathrsfs}

\newcommand{\PhiRet}{\Phi^{\rm ret}}
\newcommand{\PhiR}{\Phi^{\rm R}}
\newcommand{\PhiS}{\Phi^{\rm S}}
\newcommand{\D}{\Box}
\newcommand{\Seff}{\mathcal{S}_{\rm eff}}
\newcommand{\bPhiR}{\bar{\Phi}^{\rm R}}
\newcommand{\bPhiS}{\bar{\Phi}^{\rm S}}
\newcommand{\GS}{G^{\rm S}}
\newcommand{\NN}{\mathcal{N}_{x_0}}
\newcommand{\bNN}{\partial\mathcal{N}_{x_0}}
\newcommand{\grada}{\nabla_\alpha} 
\newcommand{\beq}{\begin{equation}}
\newcommand{\eeq}{\end{equation}}
\newcommand{\bea}{\begin{eqnarray}}
\newcommand{\eea}{\end{eqnarray}}
\newcommand{\scri}{\mathscr{I}}

\begin{document}

\title{Effective source approach to self-force calculations} 
 
\author{Ian Vega$^1$, Barry Wardell$^2$, Peter Diener$^{3,4}$}

\address{
$^1$
Department of Physics, University of Guelph, Guelph, Ontario, N1G 2W1, Canada
}
\address{
$^2$ 
Max-Planck-Institut f\"{u}r Gravitationphysik, Albert-Einstein-Institut, 14476 Potsdam, Germany 
} 
\address{
$^3$ 
Center for Computation and Technology, Louisiana State University, Baton Rouge, LA 70803, U.S.A.
}
\address{
$^4$ 
Department of Physics and Astronomy, Louisiana State University, Baton Rouge, LA 70803, U.S.A.
}
\ead{ianvega@uoguelph.ca, barry.wardell@aei.mpg.de, diener@cct.lsu.edu}
%
%
%
%
 

\begin{abstract}
Numerical evaluation of the self-force on a point particle 
is made difficult by the use of delta functions as sources. 
Recent methods for self-force calculations avoid delta functions
altogether, using  instead a finite and extended ``effective source'' for a point 
particle. 
We provide a review of the general principles underlying this strategy, 
using the specific example of a scalar point charge moving in a black 
hole spacetime. We also report on two new developments: (i) the 
construction and evaluation of an effective source for a scalar 
charge moving along a generic orbit of an arbitrary spacetime, and 
(ii) the successful implementation of hyperboloidal slicing that 
significantly improves on previous treatments of boundary conditions used 
for effective-source-based self-force calculations. Finally, 
we identify some of the key issues related to the 
effective source approach that will need to be addressed by 
future work.
\end{abstract} 

\maketitle  
 
\section{Introduction} 
\label{intro} 

One of the most important sources of low-frequency gravitational
waves for the LISA satellite are the so-called extreme-mass-ratio 
binary inspirals (EMRIs). These systems consist of a massive
black hole (such as those believed to reside in the centers of 
most galaxies) and a solar mass compact object orbiting around 
it. A typical mass ratio ($\mu/M$) for these systems would be 
around $10^{-6}$. The compact object gradually spirals into the 
massive black hole, making $O(M/\mu)$ full revolutions before 
making a final plunge that takes place in just one dynamical time 
scale.

The science potential underlying the detection and measurement of 
EMRIs by LISA is tremendous \cite{amaro-seoane-etal:07}. 
But in order to gain the most science from LISA, exquisitely accurate 
models of these astrophysical events must be readily available. 
The benchmark goal is to accurately predict the phase of EMRI 
gravitational waveforms throughout the entire inspiral. 
This remains an outstanding challenge. 

The small mass ratio reflects the huge disparity of 
physical scales characterizing EMRIs, and this disparity in turn 
is the main reason why EMRIs are such a challenge to model numerically. 
Modeling these systems requires long evolutions of over $M/\mu$ cycles, 
and very high spatial resolutions in order to account for the dynamics 
of the much smaller compact object. These conditions prove
to be prohibitive given the current reach of numerical relativity. 

On the other hand, the small mass ratio is a clear invitation for
the use of the venerable tools of black hole perturbation 
theory. From this perspective, 
the compact object can be reliably replaced by a point mass. The 
emphasis then is on how the presence of the point mass perturbs the spacetime of 
the larger massive black hole, which ultimately manifests 
in the wave zone as gravitational waves. By working with 
a point mass, an assertion is made that the internal 
degrees-of-freedom of the small compact object are negligible 
and that its bulk motion is the only relevant property for 
the purpose of waveform calculations. All this is warranted 
by the extreme mass ratio involved.

Gravitational wave emission by point masses moving around black hole spacetimes 
has been studied for a long time, starting with early calculations by Davis 
\emph{et.al.} \cite{davis-etal:71} and Detweiler \cite{detweiler:78}, 
which all rely on the seminal works of Regge and Wheeler 
\cite{regge-wheeler:57}, 
Zerilli \cite{zerilli:70}, and Teukolsky \cite{teukolsky:73}. In all of the early waveform 
calculations, the motion of the particle was assumed to be geodesic, 
so that while the perturbation on the metric is calculated, the 
back reaction of this perturbation on the motion of the particle 
was ignored. For LISA-specific data analysis requirements, such 
back reaction effects must be taken into account. Specifically, in order 
to accurately track the phase of the gravitational waveform, 
it is essential to include corrections to geodesic motion arising
from the influence of the \emph{self-force}. 

The genesis of contemporary self-force analysis can be traced 
back to derivations by Mino, Sasaki and Tanaka \cite{mino-etal:97}, 
and  Quinn and Wald \cite{quinn-wald:97}, who independently 
derived a formal expression for the self-force that is now called 
the MiSaTaQuWa equation. (An analogous expression for the scalar
charge was derived by Quinn \cite{quinn:00}.)
Since these early derivations, much of the work 
on the self-force has focused on developing practical tools for its 
evaluation. (It is notable, however, that more rigorous 
derivations of the gravitational self-force have recently been 
provided by Gralla and Wald \cite{gralla-wald:08} and 
Pound \cite{pound:10a}.)

The challenge in evaluating the self-force can be traced to the 
fact that it is formally given by an integral of the retarded 
Green function over the entire past history of the particle. Except for 
rather trivial trajectories in simple spacetimes 
(e.g. a static particle on Scwharzschild), this integral cannot 
be done analytically. Moreover, even as one resorts to 
numerical evaluation, caution must be taken in handling the 
distributional nature of the source and the singular fields 
naturally associated with point particles. All this notwithstanding,
after more than a decade of effort by various workers, a standard 
practical method of self-force evaluation has emerged. 
First proposed by Barack and Ori \cite{barack-ori:00}, it has
come to be known as the \emph{mode sum method}. This technique has 
since been implemented to compute self-forces on particles for a 
variety of cases \cite{barack:00,barack-burko:00,barack-lousto:02,burko:00c,detweiler-etal:03,diazrivera-etal:04,haas-poisson:06,barack-sago:07,haas:07,warburton-barack:10,barack-sago:10}. 

\vspace{1em}
\noindent{\em Barack-Ori mode sum method} 

The Barack-Ori mode sum method is a practical regularization 
scheme that removes unnecessary pieces from the retarded field 
of a particle, ultimately resulting in the self-force. It is 
so named because it builds upon a decomposition of the 
retarded field into spherical-harmonic modes. 

For concreteness, we shall consider the case of a minimally-coupled 
scalar charge moving in a black hole spacetime. This is a simpler toy 
problem that, from the standpoint of self-force evaluation, 
nevertheless shares much of the key technical features of the 
point mass case. The relevant field equation, in this case, 
is given by 
\begin{equation}
  \D \PhiRet = q\delta(x,x_0(\tau))
  \label{eqn:retfieldeqn}
\end{equation}
where $\D$ is the curved spacetime d'Alembertian, and $\delta(x,x_0(\tau))$
is the four-dimensional Dirac delta function representing the scalar particle 
of charge $q$, whose worldline is given by $x_0(\tau)$. The resulting retarded field, $\PhiRet$, and its gradient, $\nabla_\alpha \PhiRet$, are
singular at the location of the particle. However, the mode sum 
approach takes advantage of the fact that the spherical harmonic $l$-modes
of the gradient of the retarded field, $(\nabla_\alpha\PhiRet)_{l}$, defined through
\begin{equation}
  \nabla_\alpha\PhiRet = \sum_l^{\infty} (\nabla_\alpha\PhiRet)_l = 
  \sum_l^{\infty} \left(\sum^l_{m=-l} (\nabla_\alpha\PhiRet)_{lm}\right), 
\end{equation}
all turn out to be finite at the particle's location. 
Being finite, each of the modes can be computed numerically. 
Certain pieces are then subtracted from each of these modes 
in such a way that the remainders sum up to the correct finite 
self-force. These pieces are derived from a careful local analysis of the 
singular structure of the retarded field at the particle's 
location, and are codified in so-called \emph{regularization parameters},
$\{A_\alpha, B_\alpha, C_\alpha, \ldots \}$, which depend only 
on the background spacetime and on the orbital parameters describing the 
motion of the particle \cite{barack-ori:02,barack-ori:03a,barack-ori:03b,detweiler-etal:03,haas-poisson:06}. As such, the self-force is 
computed as 
\begin{equation}
    F_\alpha = \sum_l \left[(\nabla_\alpha \PhiRet)_l - (A_\alpha)(l+1/2) - B_\alpha - 
    \frac{C_\alpha}{l+1/2} - \ldots \right].
    \label{eqn:sfmodesum}
\end{equation}
This mode sum will converge with the use of only the 
first three regularization 
parameters $\{A_\alpha,B_\alpha,C_\alpha\}$. 

\vspace{1em}
\noindent{\em Detweiler-Whiting decomposition} 

Understanding of the  Barack-Ori mode sum method was enriched 
by Detweiler and Whiting's (DW) discovery \cite{detweiler-whiting:03} 
that the self-force on a particle can be entirely attributed to a 
(smooth) solution of the homogeneous field equation: 
\begin{eqnarray}
  \D \PhiR = 0, \\
  F_\alpha = \nabla_\alpha \PhiR, \label{eqn:gradPhiR}
\end{eqnarray}
where $\PhiR$ is known as the \emph{regular field}. 
This is akin to Dirac's \cite{dirac:38} 
classic result in the flat spacetime 
case showing that radiation reaction is due solely to the 
``half-retarded minus half-advanced'' field, which is
a smooth homogeneous solution of the wave equation.

A key point in the DW analysis was the correct identification
of the \emph{singular field}, $\PhiS$, which is the part of the retarded field that has no contribution to the self-force. 
The singular field is a solution of the same   
inhomogeneous field equation as $\PhiRet$ in the normal neighborhood 
of the particle's location:
\begin{equation} 
  \D \PhiS = q\delta(x,x_0(\tau)), \,\,\, x \in \mathcal{N}_{x_0}, 
\end{equation}
where we denote the normal neighborhood by $\mathcal{N}_{x_0}$.
As such, the singular field shares the same singularity structure 
as the retarded field, and yet it differs from the retarded 
field in that it does not depend on the 
details of the particle's distant past history. 

The regular field is then essentially the 
difference between retarded and singular fields:
\begin{equation}
\PhiR := \PhiRet - \PhiS.
\end{equation}
Therefore, from (\ref{eqn:gradPhiR}), the self-force is 
\begin{equation}
    F_\alpha = (\nabla_\alpha\PhiRet - \nabla_
    \alpha\PhiS)|_{x=x_0},
\end{equation}
from which the regularization parameters of (\ref{eqn:sfmodesum})
turn out, in effect, to be just the $l$-modes of 
$\nabla_\alpha\PhiS$ evaluated at $x_0$.

\vspace{1em}
\noindent{\em An elementary self-force calculation}

It is pedagogically useful to note that 
a self-force calculation is already encountered in any  
undergraduate course in electromagnetism.  We recall the 
problem of a point charge $q$ in the vicinity of an infinite 
grounded conducting plane. A standard problem is to compute 
the force that needs to be exerted on the charge in order to 
keep it at a fixed distance away from the plane. For concreteness, 
assume that $q$ is located at ${\bf x}_0 = (0,0,a)$, in 
Cartesian coordinates, and that the conducting plane is at $z=0$. 

This problem may of course be solved with the method of images. 
In order to maintain a zero electrostatic potential along the conducting 
plane, one imagines an image charge, $-q$, located at $-{\bf x_0}$. The 
full electric field, ${\bf E}^{\rm full}$, around $q$ will be the superposition 
of the electric fields due to the real and image charges, say ${\bf E}^{(q)}$ 
and ${\bf E}^{(-q)}$, respectively. The electric field,  ${\bf E}^{\rm full}$, 
is what is felt by an observer in the vicinity of the $q$, but it is singular 
at the location of $q$.

Now, to compute the force on the real charge, one just subtracts 
its Coulomb field from the full electric field, giving 
\begin{equation}
  {\bf F} = q({\bf E}^{\rm full} - {\bf E}^{(q)}) = q{\bf E}^{(-q)} = \frac{q^2}{4a^2}\hat{{\bf z}} \,\,\,(\rm in\,\,gaussian\,\,units),
\end{equation}
which is finite at the location of $q$. The rationale behind this subtraction 
is that ${\bf E}^{(q)}$ is isotropic around the charge $q$, and therefore cannot
exert a net force on it. 

The steps in this elementary calculation are directly analogous to more
complicated self-force calculations for the case of particles moving in a 
curved spacetime. It consists of (a) computing the full field,  ${\bf E}^{\rm full}$ 
(which is the direct analogue of $\grada \PhiRet$), (b) identifying a singular part of 
the full field that exerts no force on the charge,  ${\bf E}^{(q)}$ (\emph{cf.} $\grada \PhiS$), 
and (c) subtracting these two fields to give a smooth remainder at the location 
of the charge. What remains after this subtraction is then solely responsible for the 
self-force. For this particular problem, it is ${\bf E}^{(-q)}$ that takes the place of the 
gradient of the Detweiler-Whiting regular field. It is a solution to the 
homogeneous field equation in the vicinity of the real charge (and 
therefore smooth there). It is the force  due to this field 
that must be balanced by some external force in order to keep the 
charge in a fixed location.

\section{Self-force from an effective source}

The perspective arising from the work of Detweiler and Whiting 
is that a self-force calculation amounts to computing the smooth 
regular field, $\PhiR$, and evaluating its gradient at the particle's 
location. This idea motivates new strategies for
self-force calculation. One specific strategy enables the use
(2+1) and (3+1) grids. The ideas of this section were first independently 
presented by Barack and Golbourn \cite{barack-golbourn:07,barack-etal:07} and
Vega and Detweiler \cite{vega-detweiler:08}. Development of 
(2+1) \cite{dolan-barack:10} and (3+1) \cite{vega-etal:09} 
codes for self-force calculations has since continued to progress. 

As pointed out in the previous section, the traditional focus in a 
self-force calculation was on first solving (\ref{eqn:retfieldeqn}) for 
the retarded field. This is difficult because delta functions 
are not easily represented on a numerical grid. Moreover, it is 
not clear how to represent the retarded field on a (3+1) grid, 
given its singular nature at the particle location. Both 
issues can be circumvented by adopting the DW strategy of 
splitting the retarded field into its (smooth) regular and 
(non-smooth) singular parts. 
The idea is simple. Writing (\ref{eqn:retfieldeqn}) in terms of 
$\PhiR$ and $\PhiS$, we have 
\begin{eqnarray}
  \D (\PhiR +\PhiS) = q\delta(x,x_0)  \,\,\, x\in \NN \\
  \rightarrow \D \PhiR = 0  \,\,\, x\in \NN.  \label{eqn:Rfield}
\end{eqnarray}
Thus, given an analytic expression for $\PhiS$, 
the original field equation for 
$\PhiRet$ can be converted into a homogeneous equation
for $\PhiR$. Now this may look a bit unsettling because 
we seem to have made the source disappear entirely;
the explicit dependence of the fields on $q$ is apparently lost.
But one must not overlook the fact 
that (\ref{eqn:Rfield}) is actually valid only in the normal 
neighborhood of the particle. The singular field is 
well-defined only in the 
normal neighborhood, and therefore (\ref{eqn:Rfield}) 
is likewise restricted to only this region. 

To compute 
$\PhiR$, one needs to supplement 
(\ref{eqn:Rfield}) with boundary conditions on the boundary\footnote{Strictly speaking, the singular field is only defined inside the normal neighborhood and not its boundary. The boundary condition must then be applied just inside the normal neighborhood boundary. We therefore define $\partial\mathcal{N}_{x_0}$ to be the boundary just inside the edge of the normal neighborhood.}  
of the normal neighborhood, $\partial\mathcal{N}_{x_0}$. 
These conditions will depend on the values 
of the singular and retarded fields on $\partial\mathcal{N}_{x_0}$ 
(and thus on $q$). 
The retarded field outside the normal neighborhood 
will arise from solving the same homogeneous field equation with 
outgoing wave boundary conditions at infinity and another 
condition on its inner boundary $\partial\mathcal{N}_{x_0}$ 
enforcing consistency between $\PhiR$, $\PhiS$, and $\PhiRet$. 
Altogether then, the splitting of the retarded field turns the 
original field equation into
\begin{eqnarray}
  \D \PhiR = 0 , \,\,\, &x\in \mathcal{N}_{x_0} \\  
  \D \PhiRet = 0,  \,\,\, &x\notin \mathcal{N}_{x_0}
\end{eqnarray}
together with the matching condition $\PhiRet = \PhiR+ \PhiS$ at the 
boundary $\partial\mathcal{N}_{x_0}$, 
and outgoing wave boundary conditions at infinity. Thus, the DW splitting 
of the retarded field has turned (\ref{eqn:retfieldeqn}) 
into homogeneous field equations for $\PhiR$ and 
$\PhiRet$ in two separate computational domains. 

Without delta functions as sources, this set-up can now 
be tackled more readily on a numerical grid. 
Moreover, since one would be solving for $\PhiR$, which is smooth
at the particle, no regularization is necessary to evaluate 
the self-force; one would simply have to evaluate $\nabla_\alpha \PhiR$ 
at the location of the particle. And finally, scalar radiation waveforms are 
also just as easily computed by evaluating the retarded field
in the wave zone. 

From a practical standpoint, the use of $\partial\mathcal{N}_{x_0}$ as the 
boundary for the two distinct computational domains is bound to be inconvenient. 
Fortunately, this is not actually necessary. One is in fact permitted to choose any 
boundary so long as \emph{a} singular field can be explicitly 
computed on it and an appropriate matching condition between 
$\PhiRet$ and $\PhiR$ can be enforced. 
If one chooses a boundary outside $\NN$, then strictly speaking 
the DW singular field will no longer be defined there, and an 
extension of the singular field, call it $\bPhiS$, will be 
needed. This extended singular field must mimic the DW singular field 
close to the particle, but can be different anywhere else. 
Any extension will suffice provided that 
$\bPhiS$ and $\D \bPhiS$ are finite and 
calculable at the boundary, though ideally, one would want to choose 
a smooth extension so as not to introduce any 
unnecessary non-smoothness in the effective source. These 
considerations emphasize the point that the essential part of the 
DW singular field is its behaviour close to the particle. 
(Barack and Golbourn \cite{barack-golbourn:07} 
call these extensions of $\PhiS$ \emph{punctures}. Their 
original notion of ``puncture'' included the requirement 
that it have a straightforward $m$-mode expansion. But in 
\cite{dolan-barack:10} they have since adopted the DW singular 
field as a starting point.) 

With the freedom in choosing $\bPhiS$, it is also possible to 
completely avoid having to use two separate computational domains 
\cite{vega-detweiler:08,vega-etal:09}. 
The trick is to simply choose $\bPhiS$ to be a modulated version of 
the original singular field, such that $\bPhiS$ is forced to vanish 
outside an arbitrarily chosen neighborhood of the particle. This is 
achieved with a smooth window function, $W$, that is compactly 
supported within this neighborhood. As an example, consider the 
choice $\bPhiS := W\PhiS$, such that $W$ smoothly
transitions to zero as it approaches $\bNN$ and is zero 
everywhere outside $\NN$. The redefined regular field would then
be $\bPhiR := \PhiRet - W\PhiS$. In this case, the window function 
already forces $\bPhiR$ to equal $\PhiRet$ at $\bNN$ and 
outside $\NN$, so that no matching would have to be imposed at 
$\bNN$. The resulting field equation is then
\begin{equation}
   \D \bPhiR = q\delta(x,x_0(\tau))-\D(W\PhiS).
\end{equation}
Notice now that no distinction is drawn between the original two computational 
domains, and the only boundary conditions that need to be imposed are 
those at infinity. Outside 
the chosen neighborhood, the source on the right hand side vanishes, and so 
one is left with the (homogeneous) field equation for the retarded field. The new 
field, $\bPhiR$, thus smoothly transitions into $\PhiRet$ , and a 
single computational domain for $\bPhiR$ would suffice. 

The window function must satisfy certain conditions for the new 
regular field to retain the property that its gradient at the 
particle also equals the correct self-force, 
just as the original DW regular field. The necessary conditions which must be 
imposed on $W$ are:
\begin{enumerate}
  \item $W$ = 1 + $f$, such that $f \rightarrow O(\epsilon^n)$ and $\nabla_\alpha 
    f \rightarrow O(\epsilon^{n-1})$ as $\epsilon \rightarrow 0$; 
\item $W$ is smooth;
\item $W = 0$, $x\notin \NN$, 
\end{enumerate} 
where $\epsilon$ is some measure of distance away from the 
worldline, and $n$ is an integer that must be $\geq 3$. The lower bound in $n$
comes from the requirement that each of the last two terms on the right hand side of 
\begin{equation}
  \nabla_\alpha \bPhiR = \nabla_\alpha(\PhiRet -\PhiS) - (\nabla_\alpha f)\PhiS - f(\nabla_\alpha \PhiS) 
\end{equation}
vanish as $\epsilon \rightarrow 0$, i.e. as the particle is approached. 
Close to the particle both the singular and retarded fields behave 
like Coulomb fields in that they diverge as 
$\PhiS \sim O(\epsilon^{-1})$ and 
$\nabla_\alpha\PhiS \sim O(\epsilon^{-2})$. Condition (i) ensures 
that $\bPhiR$ does not differ from $\PhiR$ in 
any essential way as one approaches the particle, so that the gradient 
of $\bPhiR$ at the location of the charge still gives the 
correct self-force.

The field equation to be solved for $\bPhiR$ is then simply 
\begin{equation}
  \D \bPhiR = \Seff(x,x_0,u_0),
  \label{eqn:Rfieldeqn}
\end{equation}
where the \emph{effective source} is defined to be
\begin{equation}
 \Seff := q\delta(x,x_0) - \D(W\PhiS). 
  \label{eqn:Seff}
\end{equation}
This effective source is \emph{smooth}, and will  
depend on the position, $x^\alpha_0$, and four-velocity, $u^\alpha_0$, of the particle. 
The delta function of the first term is cancelled by another delta function 
coming from the second term: 
\begin{eqnarray}
  \Seff &=& q\delta(x,x_0) - \nabla_\alpha W \nabla^\alpha\PhiS  - \PhiS\Box W -W\Box \PhiS  \nonumber \\ &=& \left(q\delta(x,x_0) - W\Box\PhiS\right) - \nabla_\alpha W \nabla^\alpha\PhiS  - \PhiS\Box W  \nonumber \\ &=& -f\delta(x,x_0)  - \nabla_\alpha W \nabla^\alpha\PhiS  - \PhiS\Box W \nonumber \\ \Seff &=& - \nabla_\alpha W \nabla^\alpha\PhiS  - \PhiS\Box W \nonumber.
\end{eqnarray}
The third line follows from the fact that the singular field, $\PhiS$, 
is a local solution to the inhomogeneous field equation. The delta 
function no longer appears in the last line because $f$ 
is chosen to vanish on the worldline.

Solving for $\bPhiR$ in (\ref{eqn:Rfieldeqn}) with whatever numerical method one prefers, the 
self-force is then just 
\begin{equation}
  F_\alpha = (\nabla_\alpha \bPhiR)|_{x=x_0}. 
\end{equation}

\section{Generic effective source for a scalar point charge}

In the previous section, we expounded on the idea of evaluating 
the self-force by solving the field equations with a \emph{smooth} 
effective source, $\Seff$, as defined by (\ref{eqn:Seff}). To 
emphasize that this source does not contain a delta function, 
we may rewrite this as
\begin{equation}
  \Seff = \left\{ \begin{array}{c l} 0, & x = x_0 \\ -\Box(W\PhiS), & 
    x \ne x_0. \end{array} \right.
  \label{eqn:effsourcedefn}
\end{equation}

If a window function
is not used, $\Seff$ simplifies considerably, since it 
completely vanishes inside the normal neighborhood. As already 
indicated, a boundary around the particle would then have to be chosen 
where the matching condition between $\PhiRet$ and 
$\PhiR$ is to be imposed. These statements assume, however, that an 
\emph{exact} expression for $\PhiS$ is available, which unfortunately
is true only for the 
simplest of cases (e.g. a static particle in static black 
hole spacetimes). Generally, only an approximation to $\PhiS$ is attainable, 
and the best one can do is to write 
\begin{equation}
  \PhiS = \tilde{\Phi}^S + O(\epsilon^n), 
  \label{eqn:approxPhiS}
\end{equation}
where $\tilde{\Phi}^S$ approximates the exact singular field 
up to terms that scale as $\epsilon^n$, as $\epsilon \rightarrow 0$. 
As a result, the 
effective source will generally not vanish in the vicinity 
of the particle.

Consequently, the effective source method rests on the 
ability to write down an expression for (a non-zero) $\Seff$ 
in terms of the coordinates in which one wishes to solve 
the field equation. This shall 
be the focus of the present section. In this section, we shall 
rely on the methods of bi-tensor theory, as covered by
Poisson \cite{poissonLLR:04}.

Writing down the effective source essentially amounts to 
taking derivatives of the Detweiler-Whiting singular field 
(and the window function, if one chooses to use one). The DW 
singular field is defined through the singular Green function 
\begin{equation}
  \GS(x,x') = \frac{1}{2}U(x,x')\delta(\sigma(x,x')) 
  + \frac{1}{2}V(x,x')\theta(\sigma(x,x')), 
  \label{eqn:Gs}
\end{equation}
where $U(x,x')$ and $V(x,x')$ are bi-scalars that are regular for
$x\rightarrow x'$. This form of the Green function is valid
so long as $x$ and $x'$ are connected by a unique geodesic; that is,
in the normal neighborhoods of $x$ and $x'$.

With this, the singular field is
\begin{equation}
  \PhiS = q \int_\gamma \GS(x,z(\tau)) d\tau,
  \label{eqn:PhiSdefn}
\end{equation}
where the integration is taken over the 
wordline of the particle, $\gamma$. The singular Green 
function has support only when $\sigma(x,x') \geq 0$; 
that is, when $x$ and $x'$ are either null- or spacelike-related.
This restricts the domain of integration to a mere portion of 
the worldline. With a change of variable from $\tau$ to $\sigma$, 
one can re-express this as
\begin{equation}
  \PhiS = \frac{q}{2\sigma_{\alpha'}u^{\alpha'}}U(x,x') 
  - \frac{q}{2\sigma_{\alpha''}u^{\alpha''}}U(x,x'') 
  - \frac{q}{2}\int^v_u V(x,z(\tau)) d\tau,
  \label{eqn:PhiSsimp}
\end{equation}
where $\sigma_{\alpha'} := \sigma_{\alpha'}(x,x')|_{x'=z(u)}$, $\sigma_{\alpha'} 
:= \sigma_{\alpha''}(x,x'')|_{x''=z(v)}$, and $u$ and $v$ are the 
retarded and advanced times. Respectively, they are defined to 
be the proper times (along $\gamma$) at which a past-directed 
and future-directed null geodesics emanating from 
$x$ intersect the worldline. The retarded and advanced 
times depend implicitly on $x$ uniquely through the relations
$\sigma(x,z(u)) = 0$ and $\sigma(x,z(v))=0$. 

\subsection{Haas-Poisson expression for the singular field}
The expression for the singular field given in (\ref{eqn:PhiSsimp}) is very 
general. It is valid for any worldline in any spacetime provided the field point $x$ is sufficiently close to the worldline that the singular Green function can be defined (i.e. within the normal neighborhood). In practice, it is only in very simple spacetimes 
 that $U(x,x')$, $U(x,x'')$, $\sigma$ and $V(x,x')$ may be computed exactly. In many curved spacetimes of interest (including Schwarzschild and Kerr) this is not the case and one must find an approximation to (\ref{eqn:PhiSsimp}).

In the present work, we choose a covariant series expansion of (\ref{eqn:PhiSsimp}), taken to second order in the geodesic distance from the field point to the world line, as an approximation to $\Phi^{\rm S}$. In doing so, we use the methods described in Refs.~\cite{poissonLLR:04} and \cite{haas-poisson:06} to consolidate the dependence of $\Phi^{\rm S}$ on the advanced and retarded points $x'$ and $x''$ into a single arbitrary point $\bar{x}$ on the worldline. This has the additional advantage of making the dependence of $x'(x)$ and $x''(x)$ on $x$ explicit, so that $\bar{x}$ is truly an arbitrary point on the worldline (sufficiently close to $x'$ and $x''$) with no implicit dependence on $x$. We additionally make use of the techniques of Ref.~\cite{ottewill-wardell:10} to compute covariant expansions of all required bi-tensors.

Given the primary motivation of studying black hole spacetimes such as Schwarzshild and Kerr, we make the assumption that the spacetime is vacuum (i.e.\  $R_{ab} = 0$). In doing so, we find that near to the world-line
\begin{eqnarray}
U(x,x') \approx U(x,x'') &=& 1+ \mathcal{O}(\epsilon^4) \nonumber \\
V(x,z(\tau)) &=& \mathcal{O}(\epsilon^4).
\end{eqnarray}
For an $\mathcal{O}(\epsilon)$ effective source, we must keep terms in $U(x,x')$ and $U(x,x'')$ up to $\mathcal{O}(\epsilon^3)$ and in $V(x,z(\tau))$ up to $\mathcal{O}(\epsilon)$. We therefore find that our approximation to $\Phi_{\rm S}$ takes the simpler form
\begin{equation}
  \PhiS = \frac{q}{2\sigma_{\alpha'}u^{\alpha'}}
  - \frac{q}{2\sigma_{\alpha''}u^{\alpha''}} + \mathcal{O}(\epsilon^3).
  \label{eqn:PhiSsimpler}
\end{equation}

Next, we follow Haas and Poisson in expanding the quantities $r_- = \sigma_{\alpha'} u^{\alpha'}$ and $r_+ = -\sigma_{\alpha''} u^{\alpha''}$ about the arbitrary point $\bar{x}$:
\begin{equation}
r_{\pm} \approx  s - \frac{\bar{r}^2-s^2}{6 s} R_{u \sigma u \sigma}- \frac{\bar{r}-s}{24 s} \left[ \left(\bar{r} \pm s\right)\left(\bar{r} \mp 2s\right) R_{u \sigma u \sigma | u} - \left(\bar{r} \pm s\right) R_{u \sigma u \sigma | \sigma} \right],
\end{equation}
where $s^2 := (g^{\bar{\alpha} \bar{\beta}} + u^{\bar{\alpha}} u^{\bar{\beta}}) \sigma_{\bar{\alpha}} \sigma_{\bar{\beta}}$  (i.e.\ the projection of $\sigma_{\bar{a}}$ orthogonal to the worldline), and $\bar{r} = \sigma_{\bar{\alpha}} u^{\bar{\alpha}}$ (the projection along the worldline) and we adopt the notation of Haas and Poisson \cite{haas-poisson:06} in defining $R_{u \sigma u \sigma | \sigma} \equiv R_{\bar{\alpha} \bar{\beta} \bar{\gamma} \bar{\delta} ; \bar{\epsilon}} u^{\bar{\alpha}} \sigma^{\bar{\beta}} u^{\bar{\gamma}} \sigma^{\bar{\epsilon}} \sigma^{\bar{\delta}}$. 
Substituting these expansions into (\ref{eqn:PhiSsimpler}), we get
\begin{eqnarray}
\label{eqn:PhiScov}
\Phi^{\rm S} &=& q\Bigg\{\frac{1}{s} + \bigg[\frac{\bar{r}^2-s^2}{6 s^3} R_{u \sigma u \sigma}\bigg] \nonumber \\
&&\quad + \bigg[ \frac{1}{24 s^3} \big( \left(\bar{r}^2 - 3 s^2\right) \bar{r} R_{u \sigma u \sigma | u} - \left(\bar{r}^2-s^2\right) R_{u \sigma u \sigma | \sigma} \big) \bigg] + \mathcal{O}(\epsilon^3)\Bigg\}.
\end{eqnarray}
Letting $\epsilon$ be a measure of the geodesic distance from $x$ to the world-line (i.e.\ $\bar{x}$), the first term here is $\mathcal{O}(\epsilon^{-1})$, the second group of terms is $\mathcal{O}(\epsilon^1)$ and the third group of terms is $\mathcal{O}(\epsilon^2)$. The error in this approximation to $\Phi_{\rm S}$  is then $\mathcal{O}(\epsilon^3)$.

The final step is to convert this covariant expression for the singular field to an expression in terms of the coordinates of the background spacetime. This is easily achieved by using the methods of \cite{haas-poisson:06} or \cite{ottewill-wardell:09} to rewrite $\sigma^{\bar{a}}$ as a coordinate expansion and then substituting the result into (\ref{eqn:PhiScov}).

\begin{figure}
\begin{center}
\includegraphics[width=7.5cm]{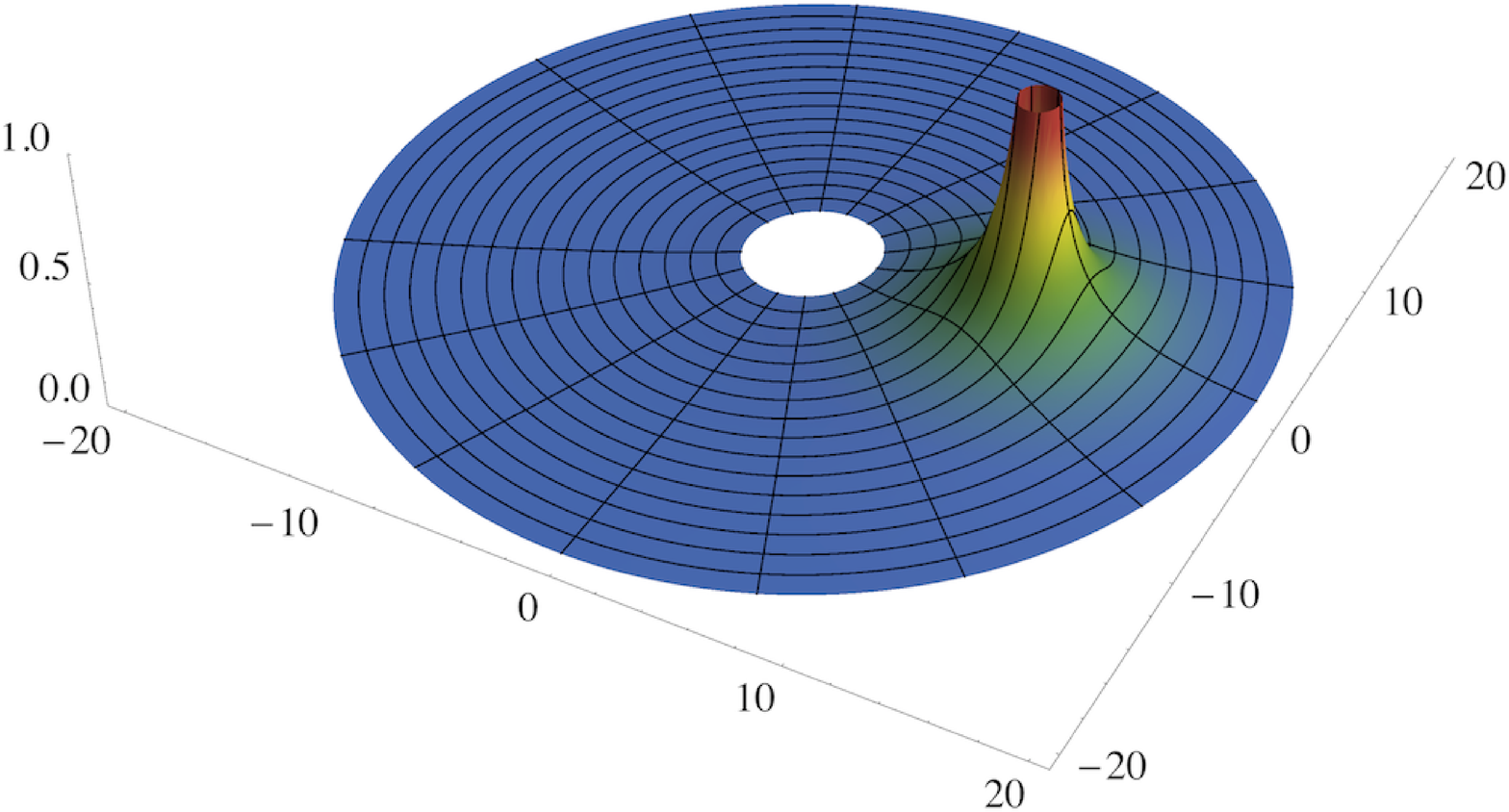}
\includegraphics[width=7.5cm]{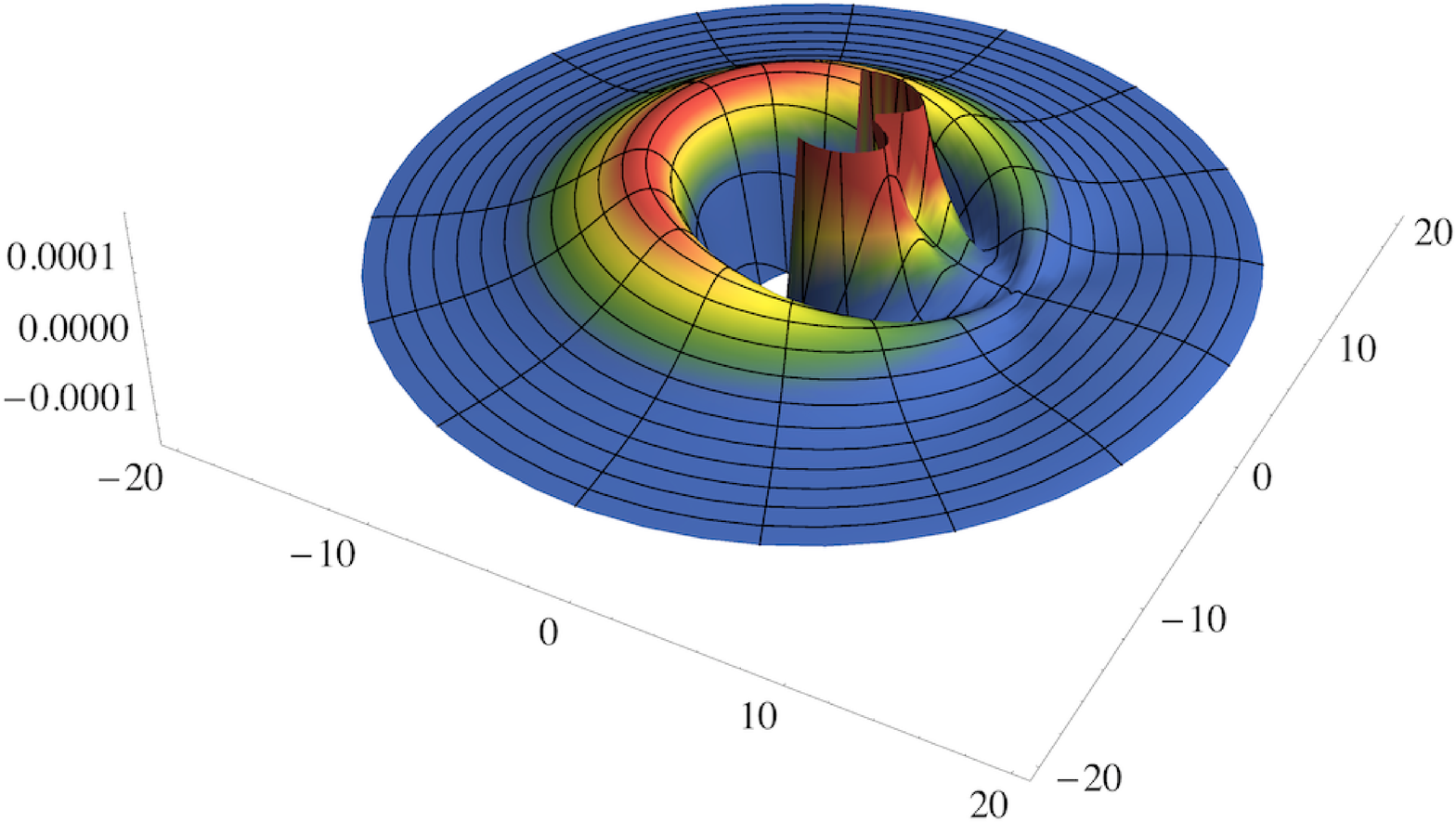}
\caption{(Color online) Approximation to the singular field (left) and corresponding effective source (right) along the equatorial plane for a particle in a circular orbit (at $r=10M$) around a Schwarzschild black hole.}
\label{fig:schw}
\end{center}
\end{figure}

With the coordinate expression for the approximate singular field at hand, what remains to be done in order to get $\Seff$ 
is to apply the d'Alembertian on the singular field. In Fig.~\ref{fig:schw} we give plots demonstrating the result of this calculation for the case of a particle in a circular equatorial orbit (at $r=10M$) around a Schwarzschild black hole. The surface plots give the value of the singular field and effective source along the equatorial plane. The shown approximation to the singular field differs from the exact singular field at $\mathcal{O}(\epsilon^3)$ and hence the effective source differs from the exact effective source at $\mathcal{O}(\epsilon)$. As there was no window function (equivalently $W=1$) used in generating these plots, the exact effective source would be $0$ and so the approximate effective source is $S_{\rm eff}= 0 + \mathcal{O}(\epsilon)$. In this plot, the particle is at the point where there appears to be a pinch in the curve.

We emphasize that the resulting effective source is for a scalar charge moving along a \emph{generic} geodesic,
and that this is the first time an effective source of this sort has appeared in print. 

\subsection{THZ coordinates}

Another route towards the effective source makes use of a 
convenient set of locally-inertial coordinates first introduced
by Thorne and Hartle \cite{thorne-hartle:85} and Zhang \cite{zhang:86}. 
This would entail directly performing the integration in 
(\ref{eqn:PhiSdefn}) as expressed in these coordinates. It was 
through this approach that the first coordinate expression for 
the scalar singular field was derived (albeit for the restricted 
case of a circular orbit in Schwarzschild) \cite{detweiler-etal:03}. 

The appeal of this coordinate frame, in the self-force context, is due 
to the fact that they result in the background metric and d'Alembertian 
taking on particularly  convenient forms. As a result, the singular field can 
be expressed simply as 
\begin{equation}
  \PhiS = \frac{q}{\sqrt{X^2+Y^2+Z^2}} + O(\epsilon^3)
  \label{eqn:thzSfield}
\end{equation}
in THZ coordinates $\{T,X,Y,Z\}$.

In this form, the complexity of the singular field is hidden in the transformation from the coordinates
of the background spacetime, $\{x\}$, to the
THZ coordinates, $\{X(x),Y(x),Z(x)\}$. This transformation was derived explicitly for the case of a 
circular orbit in the Schwarzschild spacetime \cite{detweiler-etal:03}. Progress
is being made towards deriving this transformation for a generic geodesic in an 
arbitrary spacetime.

\section{Issues in evaluating the effective source}

\noindent{\em Computational cost} 

Writing out the coordinate expression for $\Seff$ is straightforward enough to do with computer algebra 
software such as \emph{Mathematica} or \emph{Maple}, but for the order of the approximation that is sought 
here (i.e. $\Seff \sim O(\epsilon)$ as $\epsilon \rightarrow 0$), what results is a 
massively long expression whose evaluation is potentially very costly in an application 
requiring the effective source to change \emph{dynamically} in a self-consistent (3+1) simulation. 
Going from Schwarzschild to the more complicated Kerr geometry further exacerbates this issue. 

Given such considerations, computing the d'Alembertian of $W\PhiS$ with \emph{numerical} derivatives
(instead of evaluating a full analytical expression for the derivatives) becomes worth exploring. Done naively, this 
strategy is likely going to sacrifice some accuracy, but it is quite possible that this loss of accuracy will be tolerable. We can be optimistic that this is the case given the fact that 
the d'Alembertian of $\bPhiS$ can be evaluated by numerically computing derivatives of low-order polynomials. 
One can see this from (\ref{eqn:PhiScov}), which can be 
rewritten as
\begin{equation}
  q^{-1}\bPhiS = \frac{K}{S^{3/2}}, 
\end{equation}
where
  \begin{eqnarray}
  K &:= s^2 &+ \frac{1}{6} (\bar{r}^2-s^2) R_{u\sigma u \sigma} + \frac{1}{24}(\bar{r}^2-3s^2)\bar{r}R_{u\sigma u\sigma|u}  \nonumber 
  \\ &&-\frac{1}{24}(\bar{r}^2-s^2)R_{u\sigma u\sigma|\sigma}
  \\ S &:=& s^2
\end{eqnarray}
Both $K$ and $S$ when written in, say, Schwarzschild 
coordinates are just low-order polynomials 
\begin{eqnarray}
  K &=& \sum\limits_{i,j,k,l=2\atop i+j+k+l=5} A_{ijkl}(t-T)^i(r-R)^j(\theta-\Theta)^k(\phi-\Phi)^l \label{eqn:K}
  \\ S &=& \sum\limits_{i,j,k,l=2\atop i+j+k+l=5}  B_{ijkl}(t-T)^i(r-R)^j(\theta-\Theta)^k(\phi-\Phi)^l \label{eqn:S},
\end{eqnarray}
with $\{T,R,\Theta,\Phi\}$ denoting the particle's position in these coordinates. 
Numerical derivatives of these 5th-order polynomials can be computed quite accurately and cheaply. 

Exploring how best to compute the effective source with numerical derivatives is left to future work.



\vspace{1em}
\noindent{\em Catastrophic cancellation near the worldline}

Another serious issue encountered is the potentially severe round-off error incurred when evaluating 
the effective source close to the particle. This is an issue regardless of one's 
strategy for computing derivatives. It is easy to understand why
this would happen.

The approximate singular field possesses the familiar $O(\epsilon^{-1})$ Coulombic 
divergence close to the particle. Consequently, the d'Alembertian acting on $\bPhiS$ 
results in dominant terms that each scale as $O(\epsilon^{-3})$. By construction, however, the effective 
source ought to scale as $O(\epsilon)$ as $\epsilon \rightarrow 0$. This means that all the 
$O(\epsilon^{-3})$-terms should somehow cancel to give the $O(\epsilon)$ over-all behaviour of the effective source. 
This is a typical instance of catastrophic cancellation that can lead to massive round-off errors. 

For moderately small $\epsilon$, say $\epsilon = 0.1$, the subtraction of terms results in 
a number $10^{-4}$ times smaller than the magnitude of the terms themselves. This implies a loss of 
4 significant digits in the result, which may perhaps still be acceptable (assuming, say, double precision of 
$\sim$15 digits). But closer to the particle, at say $\epsilon = 10^{-3}$, one instead loses 12 significant 
digits. It is clear then that as one approaches the particle, the evaluation of the effective 
source becomes increasingly inaccurate. In \cite{vega-etal:09}, 
a very crude ``fix'' was employed to avoid this concern: whenever an evaluation very close to the particle 
was needed, a different point slightly farther from the particle was chosen in evaluating the 
effective source. Remarkably (and rather surprisingly, in retrospect), this cheap fix alone lead 
to results with errors of $\lesssim 1\%$. 

The ideal solution to this would be to get the $O(\epsilon^{-3})$-terms in the effective source to cancel 
\emph{analytically}, leaving only a remainder that would then \emph{explicitly} scale as $O(\epsilon)$.
Unfortunately, the expressions we have worked with turn out to be sufficiently complicated to prohibit this this being done in any
obvious manner. (Experiments with analytical expressions in \emph{Mathematica} give us indications that there may be some way 
of achieving this, but we have not explored this sufficiently to be able to make any concrete statements.)

Another solution to this would be to work towards a covariant expansion of the effective source. Instead 
of taking (partial) derivatives of the singular field, one can postpone the 
introduction of coordinates and find an expansion of $\Seff$ itself in powers of $\sigma^{\bar{\alpha}}$. This 
provides a clear route for cancelling the divergent terms analytically, but it comes at the expense of introducing 
errors in the effective source that grow away from the particle. However, if one restricts the use of the covariant expansion to a region very close to the
particle ($\epsilon \ll 1$), then these errors are also going to be small and potentially 
acceptable. With a covariant expression at hand, one possible strategy would be to use it for evaluations close to the particle, and to use the usual expression (i.e., computed with partial derivatives) when 
one is further away. Work by Ottewill and Wardell 
\cite{ottewill-wardell:10} has now pushed Hadamard expansions to very high order, allowing for ever more accurate approximations to the singular
field. The techniques they developed could be used to derive a high-order covariant effective source 
that would have a wider region of applicability around the particle. 

\begin{figure}
 \begin{center}
  \includegraphics[width=7.5cm]{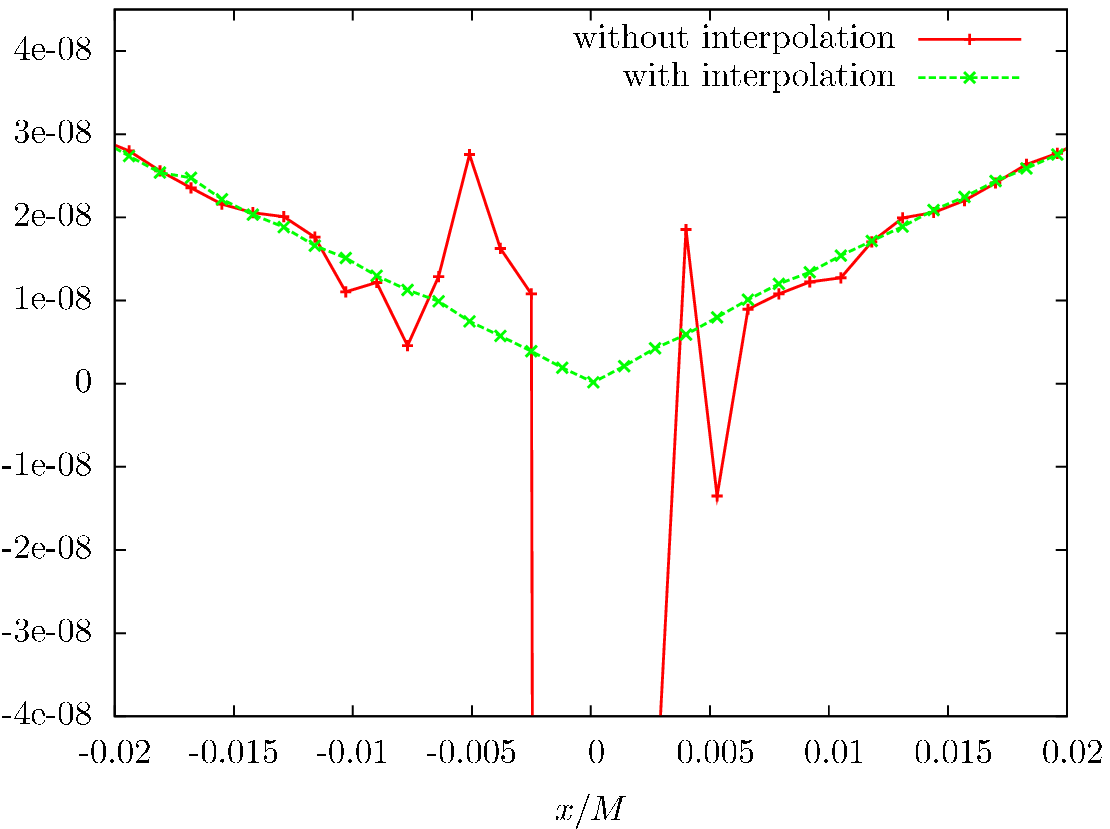}
  \includegraphics[width=7.5cm]{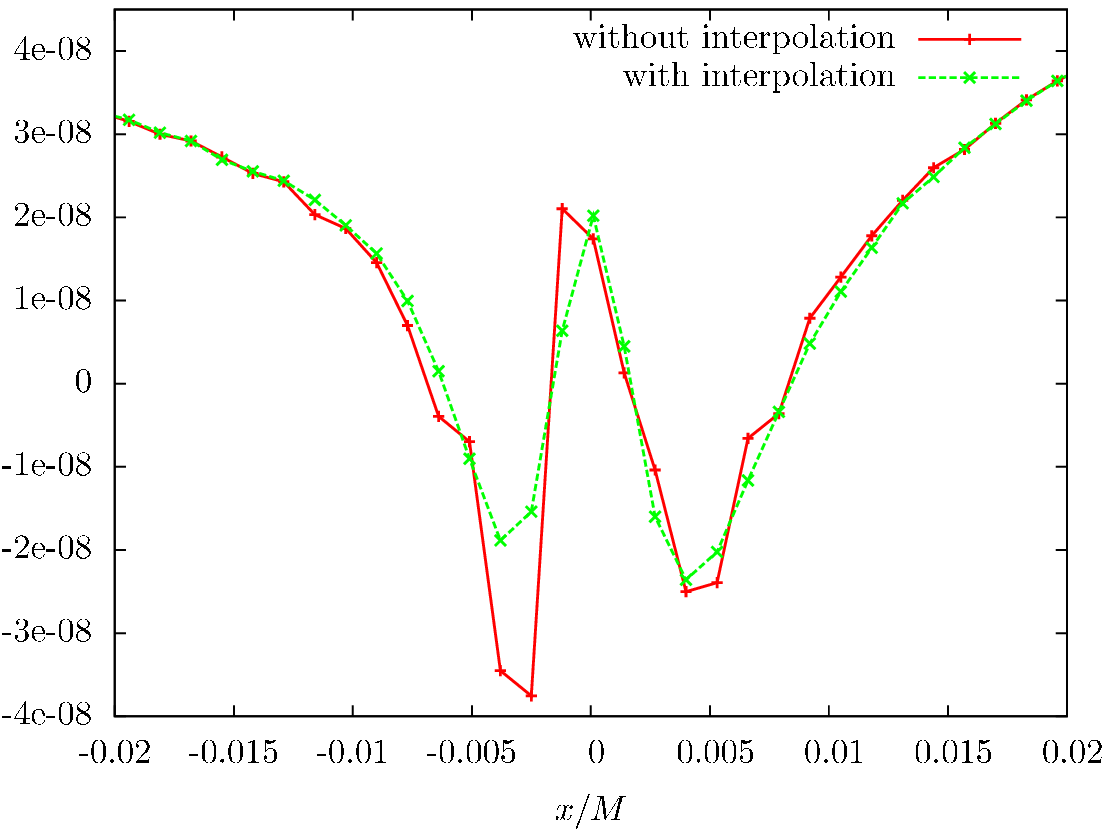}
 \end{center}
 \caption{(Color online) Interpolation of the effective source. These figures show how a simple interpolation scheme mitigates the round-off errors due to the catastrophic cancellation that occurs in the region very close to the particle. Both figures are plots of $\Seff(x,y,z=0)$, with $y/M=10M$ and $y/M=10.005M$ (in Kerr-Schild coordinates), respectively, for a scalar charge moving in a circular orbit of Schwarzschild at $r=10M$, where $r$ is the Schwarzschild radial coordinate. The scalar charge is located at $(x=0, y=10M, z=0)$. The plot on the left clearly shows the $\mathcal{C}^0$ nature of the effective source at the particle's location.} 
 \label{fig:interp}
 \end{figure}

Finally, if all else fails, a simple polynomial interpolation (see Fig. \ref{fig:interp}) 
might be the best fix for this problem. 
One would identify a small neighborhood around the particle within which round-off reaches unacceptable levels. When the effective source needs to be evaluated at a point inside this region, one would interpolate using effective source values outside the region and the known zero value of the effective 
source at the location of the particle.

Another immediate goal is to determine an optimal solution to this round-off issue, which is likely going to 
be a hybrid of the ideas presented here. 

\section{Issues for the effective source approach to self-force calculations}
The numerical experiments performed in~\cite{vega-etal:09} showed the existence
of some serious issues that must be overcome before simulations
can be performed for long enough and with high enough accuracy to reliably
extract the self-force.

The first problem is the presence of non-physical
outer boundaries arising from truncating the numerical domain at a finite
radius.  In~\cite{vega-etal:09} both codes enforced zero incoming modes at the
outer boundary.  Such boundary conditions are typically used for numerical
convenience in order to maintain stability, but are clearly unphysical since
the tail from the spacetime outside of the numerical domain is ignored.  This
essentially leads to a non-physical reflection from the outer boundary that 
will affect the extraction of the self-force as soon as the location of the
particle comes into contact with the reflected wave.  A simple remedy is of
course to push the outer boundary to larger radius, practically delaying the
influence of the outer boundary condition.  However, this comes at a steep
price in computational cost.

The second problem comes from the $C^0$ nature of the source, leading to the limited $C^2$ regularity
 of the scalar field at the particle location.  In order to obtain
the expected convergence with either high order finite differencing or spectral
methods it is required that the fields being evolved are sufficiently smooth
(otherwise the error estimates based on Taylor expansions do not hold).
With $C^2$ fields, the extracted self-force only converges to 2nd order using
finite differencing methods and between 4th and 5th order using spectral
methods.

\vspace{1em}
\noindent{\em Boundary conditions for long-term evolutions} 

The outer boundary problem has been solved using the approach described in
\cite{Zenginoglu:2009hd}.  The main idea is to use hyperboloidal slices that
approach future null infinity ($\scri^+$) at a finite coordinate radius.
Our starting point is the black hole metric $\tilde{g}$ in Kerr-Schild form
modified by a spatial coordinate transformation in order for the black hole
horizon to be a coordinate sphere even in the rotating case.  This form of the
metric is particularly well suited to our 3D multi-block code where we have a
spherical inner excision boundary (inside the black hole horizon) as well as a
spherical outer boundary.

To proceed, we introduce a new time coordinate $\tau$ related to the standard
Kerr-Schild time coordinate $t$ by
\begin{equation}
\tau = t - h(r),
\end{equation}
where $h(r)$ is called the height function.  In addition we compactify the
radial coordinate
\begin{equation}
r=\frac{\rho}{\Omega},\mbox{\hspace{2em}with\hspace{2em}}\Omega=\Omega(\rho).
\end{equation}
To remedy the fact that this leads to the metric $\tilde{g}$ being singular at
$\Omega=0$ we additionally perform a conformal rescaling to obtain
$g = \Omega^2 \tilde{g}$.  Under such a conformal rescaling the scalar wave
equation satisfies the following conformal transformation rule
\begin{equation}
\left (\Box - \frac{1}{6} R\right ) \phi = 
\Omega^{-3}\left (\tilde{\Box}-\frac{1}{6}\tilde{R}\right )\tilde{\phi}.
\end{equation}
Here $\Box$ and $R$ are the wave operator and Ricci scalar associated with the
rescaled metric $g$, while the quantities with tildes are associated with the
physical metric. The scalar fields $\phi$ and $\tilde{\phi}$ are related
through $\phi = \tilde{\phi}/\Omega$. Note that although $\tilde{R}$ is zero
(since it is a vacuum spacetime) the conformally rescaled $R$ is non-zero.

In regions where the wave equation is homogeneous (i.e.\ where $\Seff$ is zero)
we then have to solve
\begin{equation}
\Box\phi - \frac{1}{6}R\phi = 0.
\end{equation}
In regions where $\Seff$ is non-zero, we want to keep things simple and use
standard spatial slices, i.e.\ we want to transition from standard spatial
slices near the black hole and source to hyperboloidal slices in the exterior.
We do this by suitable choices for $\Omega(\rho)$ and $H=\rmd h/\rmd r$
\begin{equation}
\Omega(\rho) = \left \{\begin{array}{lcl}
                  1 & \mbox{for} & \rho\le\rho_{\mathrm{int}} \\
                  1-f + (1-\rho/S) f
                  & \mbox{for} & \rho_{\mathrm{int}}<\rho<\rho_{\mathrm{ext}} \\
                  1-\rho/S & \mbox{for} & \rho\ge\rho_{\mathrm{ext}}
               \end{array}\!\!, \right.
\end{equation}
\begin{equation}
H(\rho) = dh/dr = \left \{\begin{array}{lcl}
                  0 & \mbox{for} & \rho\le\rho_{\mathrm{int}} \\
                  \left (1 +\frac{4 M\Omega}{\rho}+\frac{(8 M^2-C^2)\Omega^2}{\rho^2} \right ) f
                  & \mbox{for} & \rho_{\mathrm{int}}<\rho<\rho_{\mathrm{ext}} \\
                  1 +\frac{4 M\Omega}{\rho}+\frac{(8 M^2-C^2)\Omega^2}{\rho^2} & \mbox{for} & \rho\ge\rho_{\mathrm{ext}}
               \end{array}\!\!. \right.
\end{equation}
Here $S$ determines the coordinate location of $\scri^+$, $M$ is the mass of
the black hole, $C$ is a free parameter and $f$ is a function that smoothly
transitions from 1 at $\rho_{\mathrm{int}}$ to 0 at $\rho_{\mathrm{ext}}$.

With this choice, the coordinate speed of in- and outgoing characteristics at
$\rho=S$ are
\begin{equation}
c_- = 0, \mbox{\hspace{2em}} c_+ = S^2/C^2.
\end{equation}
By choosing $C=S$ we can fix the outgoing characteristic speed to $c_+=1$ 
(much larger values would severely limit the size of time steps we could use).
Since $c_-=0$ we have no incoming modes (as expected at $\scri^+$) and we then
do not have to apply a boundary condition at the outer boundary.  This is
similar to the situation at the inner boundary, where the presence of a
black hole horizon ensures that all modes leave the computational domain.  We
are therefore in the remarkable situation that we do not have to apply any
boundary conditions on either the inner or the outer boundary of the 
computational domain.  It is also worth mentioning that an outgoing wave will
reach $\scri^+$ in finite coordinate time.
\begin{center}
\begin{figure}
\includegraphics[width=7.8cm]{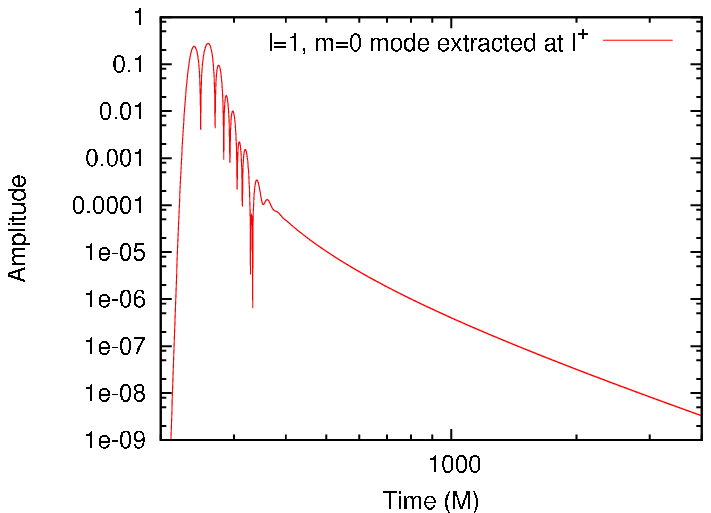}
\includegraphics[width=7.8cm]{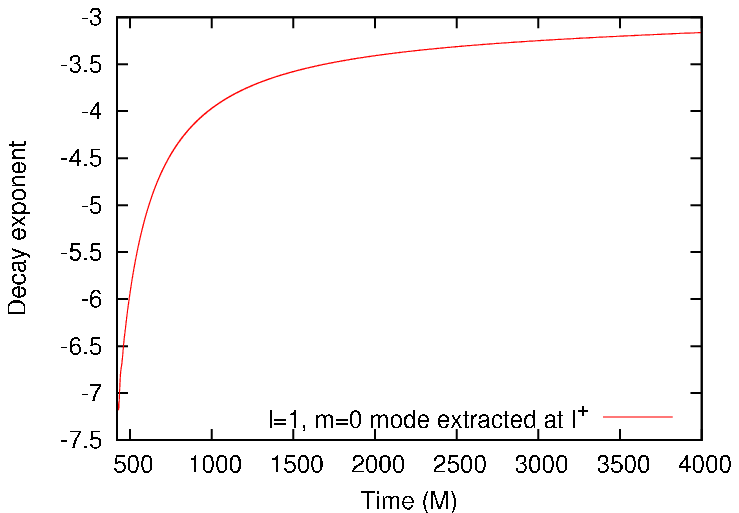}
\caption{(Color online) The left plot shows the $\ell=1$, $m=0$ mode of the
scalar field extracted at $\scri^+$ when evolving $\ell=1$, $m=1$ initial data
in a Schwarzschild background. The right plot shows the decay exponent of the
tail as a function of time.}
\label{fig:tail}
\end{figure}
\end{center}

As a first test of the performance of the new outer boundary condition we
evolved $\ell=1$, $m=0$ scalar initial data (with no source) on a Schwarzschild
($M$=1) background with $S=100 M$ and $C=100 M$.  The results presented in
Figure~\ref{fig:tail} are based on a simulation with radial resolution 
$\Delta \rho=M/15$ and 30 points per patch in the angular directions and using
8th order finite differencing.  The inner excision boundary was placed at
$\rho = 1.8M$, while the transition from spatial slices to hyperboloidal slices
occurred between $\rho=20M$ and $\rho=60M$.  The left
plot shows the amplitude of the $\ell=1$, $m=0$ mode extracted at $\scri^+$
($\rho = S = 100M$) in a log-log plot.  The wave first reaches $\scri^+$ at
around $230M$ and we then see quasinormal mode ringdown with a transition
into the tail regime.  The right plot shows the numerically extracted exponent
of the tail decay.  At late times it is expected that the tail of a $\ell=1$
mode decays as a power law $\propto t^{-3}$ at $\scri^+$ (see e.g.\ 
\cite{Hod:1999rx}).  Note that this is different from the expectation at
timelike infinity $I^+$ where the tail would decay as a power law with
exponent $-(2\ell+3)=-5$.  As can be seen from the plot, we are approaching
the value $-3$ for the decay exponent at late times.  Note that the total
evolution time is $4000 M$, which is many times more than the crossing time,
and there are no visible effects from either the inner or the outer boundary.
This is a very strict test, since it only takes minor numerical errors or
inconsistencies to significantly affect the tail behavior.

As a second test we repeated the simulation in~\cite{vega-etal:09}, that is
a scalar point charge on a circular orbit of radius $10M$ in Schwarzschild spacetime.  The simulation 
in~\cite{vega-etal:09} was performed with the outer boundary located at
$r=600M$, while the new simulation was performed with $\scri^+$ located at
a coordinate radius of $\rho=100M$.  This means that the computational cost
of the new simulation was about a factor of 6 smaller than the old one.  The
angular resolution was kept the same (40 points per patch).  In the new run
the source was turned on smoothly over one orbit.  This was done to avoid an
issue with under-resolving the spurious wave generated by turning on the
source instantaneously.  In the old simulation the coordinate speed of the
outgoing characteristics is a monotonically increasing function (approaching 1
as the coordinate radius $r\rightarrow\infty$), while in the new simulation,
due to the transition from
spatial to hyperboloidal slices, the coordinate speed drops to a minimum of
about 0.2 in the transition region and then increases to 1 at $\rho=S$.  This
has the effect that any outgoing wave experiences a compression by about a
factor of 5 as it travels through the transition region.  Turning on the source
slowly ensures that the spurious wave generated initially has long enough
wavelength that it is never under-resolved.
\begin{center}
\begin{figure}
\includegraphics[width=7.8cm]{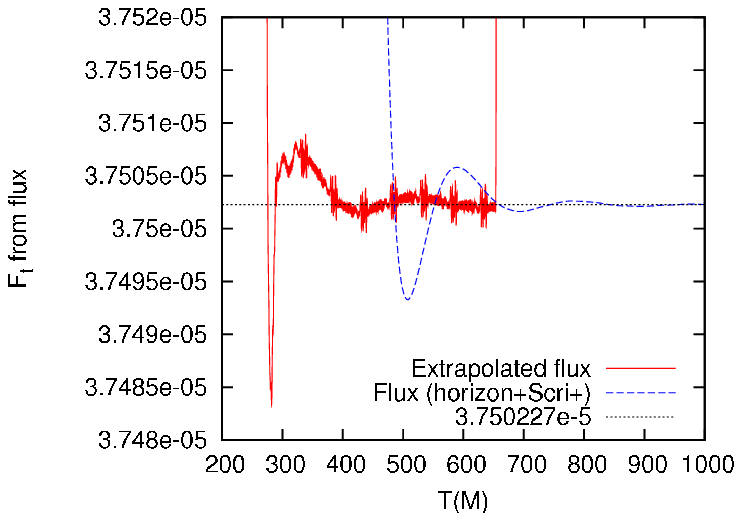}
\includegraphics[width=7.8cm]{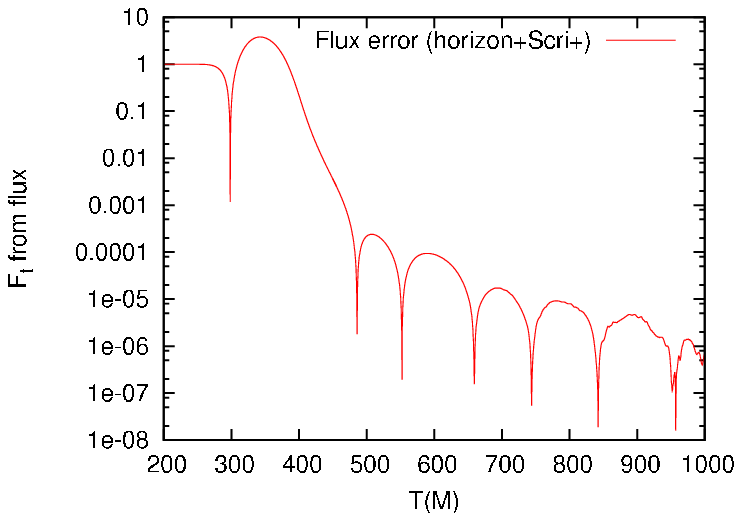}
\caption{(Color online) The left plot shows a comparison of the extraction of
the flux done at finite radii and Richardson extrapolated to spatial infinity 
(red solid line) with the extraction of the flux at $\scri^+$ (blue dashed
line).  The black dotted line, shows the result from highly accurate results
from a frequency domain code. The right plot shows a log plot of the relative
error in the extraction of the flux at $\scri^+$, measured against the
frequency domain result.}
\label{fig:flux}
\end{figure}
\end{center}

The comparison of the best extracted fluxes are shown in the left plot of
Figure~\ref{fig:flux}.  The fluxes shown are the sum of the fluxes going into
the black hole and the outgoing fluxes at infinity converted to the
corresponding time component of the self-force.  Since, in a time domain
calculation, the source is turned on at some point in time, it is expected
that it takes some time to reach a helically symmetric state and thus the
flux will start out being completely wrong (i.e. zero) and only reach the
correct value asymptotically. This can bee seen for both the solid (red) and
dashed (blue) curves.  The solid (red) curve is produced by extracting
the flux at several radii (from $50M$ to $300M$) and then Richardson
extrapolated (taking care to time shift accordingly to the light travel time
from detector to detector) to spatial infinity.  As a result there is some
high frequency noise.  In addition the curve terminates at $T=650M$ when the
outermost detector comes into causal contact with the initial spurious pulse
hitting the outer boundary.  In order to extend the curve, we would have to
redo the simulation with the outer boundary even further out.  In contrast,
the dashed (blue) curve is very smooth and shows no boundary effects for the
duration of the simulation, even with the outer boundary located at 
$\rho=100M$. This is expected to continue for as long as we would care to
extend the simulation time.  The right plot in Figure~\ref{fig:flux} shows
the relative error in the flux extraction as a function of time.  After about
5 orbits we have gotten rid of so much of the spurious waves that the relative
error is as low as 1 part in a million.  This is a quite remarkable accuracy
considering this is a full 3D simulation.

\vspace{1em}
\noindent{\em Over-all accuracy} 

As mentioned above, the accuracy in evaluating the self-force at the location
of the particle is limited by the $C^0$ nature of the effective source $\Seff$.
The effect of the non-smoothness of the source on the evaluation of the
self-force manifests itself as high frequency noise (the frequency being
determined by the time it takes the particle to move from a gridpoint to a
neighbouring gridpoint) that only converges away at 2nd order regardless of
the finite differencing order used (see~\cite{vega-etal:09} for more details).

A possible remedy is to modify the finite differencing stencil near the
particle location so that the stencil becomes more and more one sided as the
particle location is approached in order to always use smooth data to 
approximate derivatives.  However, experiments in 1D have shown that a
straightforward implementation of such adjustable finite differencing stencils
leads to an unstable numerical scheme, where high frequency noise develops and
blows up on a very short timescale.  We have not yet finished exploring 
possible ways of stabilizing such a scheme, but it is certainly possible that
such an approach in the general case is doomed to fail.  On a positive note,
we have been able to find a stable scheme for the case in 1D where the particle
is always located exactly at a grid point.  We may be able to take advantage
of that in our multiblock infrastructure by covering the neighbourhood of the
particle with an extra block that is co-moving with the particle.  This co-moving
block would be completely contained within a non-moving block.  The
communication between the blocks would be done with interpolation away from
the particle location where all fields are smooth.  Such an approach would not
be possible in the multiblock infrastructure used in~\cite{vega-etal:09} but
could be implemented in the multiblock infrastructure used 
in~\cite{Pollney:2009ut} and described in more detail in~\cite{Pollney:2009yz}.
It is not yet clear how much infrastructure work such an approach would require.

Finally, an alternative possible remedy is to construct a smoother 
effective source at the analytical level. This is now possible through 
the methods of $\cite{ottewill-wardell:10}$, and would result from
the inclusion of more terms in the covariant expression for the 
singular field, as displayed in (\ref{eqn:PhiScov}). 
This would make the coordinate expression for $\Seff$ even 
more complicated than it already is, but this increase in 
complexity could be mitigated if evaluation of the 
effective source with numerical derivatives is found adequate. 
In any case, with a $C^2$ effective
source, the amplitude of the noise would not only be decreased 
significantly compared to a $C^0$ source at the lowest resolution 
currently used, the convergence would also be 4th order instead 
of 2nd order, so that small increases in resolution would lead to 
significantly lower 
errors.

\section{Conclusion and future prospects} 

In this manuscript, we reviewed the foundations underlying the effective 
source method for self-force calculations. The approach is powerful because it 
avoids singular sources and fields, thereby obviating the need for any sort of 
post-processing regularization such as the one performed in 
the Barack-Ori mode sum method. Moreover, the effective source approach 
is, in principle, indifferent to the background spacetime, and therefore 
allows for the evaluation of the self-force on a particle
moving on a generic geodesic of an arbitrary curved spacetime. 

Not surprisingly, however, the new method comes with its attendant challenges, 
and we have outlined some of the strategies that may be pursued to overcome them.  

Throughout much of the paper, the effective source approach was presented 
mainly as a technique for evaluating the self-force for a prescribed 
geodesic orbit in a black hole spacetime. This was done to avoid (as much 
as possible) having to commit to a specific type of numerical grid 
(i.e. whether 2+1 or 3+1), as it is important to stress that the use of 
an effective source is conceptually distinct from this commitment. (Even the use of 
hyperboloidal slicing, while implemented and tested on a 3+1 grid for this 
paper, would be beneficial to 2+1 codes as well).

However, one important aspect of the effective source approach is
that by avoiding the necessity of any post-processing altogether (e.g. 
regularization or mode sums), it facilitates the transition from mere 
self-force computations to the actual self-consistent simulation of the 
dynamics of both the particle and the field. In the scalar case, all 
that is necessary is to supplement the field equation for $\bPhiR$ with 
the equation of motion of 
the particle:
\begin{eqnarray}
  \D \PhiR = \Seff(x,x_{\rm p},u_{\rm p}) \\
  \frac{D u^\alpha_{\rm p}}{d\tau} = g^{\alpha \beta}(\nabla_\beta \PhiR)|_{x=x_{\rm p}}. 
  \label{eqn:selfconsistent}
\end{eqnarray}
This set of equations can be simultaneously solved at every timestep on a (3+1) grid. 
Effective source implementations on (2+1) \cite{barack-golbourn:07} or (1+1) \cite{vega-detweiler:08} 
grids also avoid the need for regularization, but require a post-processing 
mode sum in order to compute the self-force. In these implementations, one first 
has to decompose $\Seff$ into its spherical- or spheroidal-harmonic components, which 
are then used as the sources to the corresponding reduced wave equations. Such mode-sum 
implementations stand to benefit from the inherent symmetries of the Kerr spacetime, 
and as a result have been demonstrated (in the Schwarzschild case) to 
lead to more accurate self-force results \cite{dolan-barack:10}. 

A self-consistent evolution of particle-field dynamics has thus 
far eluded the self-force community, and it would therefore be interesting 
to see if any suprises arise from it even for the case of the scalar 
charge. Particularly intriguing are the recently discovered Flanagan-Hinderer 
resonances \cite{flanagan-hinderer:10} that show up only for non-equatorial 
orbits in Kerr. With an effective source, it ought to be possible to probe the 
self-consistent behavior of a scalar charge as it goes through one of these resonances.

A generalization of the effective source approach to the gravitational case is also 
within reach. All that would be required is the construction of the corresponding 
effective point mass (in Lorenz gauge), and much of this follows from the 
methods we have described here. However, the task of then meaningfully involving this 
effective point mass in a self-consistent scheme will require some thought. 
The situation in the gravitational case is more complicated than in the scalar 
case because the former involves an unusual ``gauge condition'' (strictly, a 
controlled \emph{violation} of the Lorenz gauge) that must also be satisfied 
at every time step \cite{gralla-wald:08}. It is unclear at the moment how to 
translate this requirement into a practical framework. 

In this paper, we described notable progress on two fronts, mainly (a) the 
construction of an effective source for a scalar 
point charge moving along a \emph{generic} geodesic, and (b) 
the implementation of hyperboloidal 
slicing to resolve the issue of boundary reflections, which is 
critical for the long-term evolutions required by the effective 
source approach. We anticipate that future work shall focus on making much needed 
refinements to our current effective source, calculating the 
self-force on scalar charges moving on generic geodesics in Kerr, 
and finally, studying the self-consistent dynamics of a scalar particle
and its field.

\vspace{1em}
\noindent{\bf Acknowledgements}

\vspace{1em} \noindent We gratefully acknowledge Steve Detweiler and Eric
Poisson for valuable discussions.   We are also thankful to Sam Dolan, Leor
Barack, Jos\'e Luis Jaramillo, Michael Jasiulek and Erik Schnetter for
providing  insightful suggestions.  Portions of this research were conducted
with high performance computational resources provided by the Louisiana
Optical Network Initiative (http://www.loni.org/).  In addition this research
was supported in part by the National Science Foundation through TeraGrid
resources provided by LSU/NCSA/NICS under grant number TG-MCA02N014.  Finally,
we would like to express our gratitude to the attendees of the 2010 Capra
meeting for many illuminating conversations and to the Perimeter Institute for
providing a fruitful environment for the meeting.

\section*{References}


\begin{thebibliography}{10}

\bibitem{amaro-seoane-etal:07}
P.~Amaro-Seoane, J.R. Gair, M.~Freitag, M.C. Miller, I.~Mandel, C.J. Cutler,
  and S.~Babak.
\newblock {Intermediate and extreme mass-ratio inspirals -- astrophysics,
  science applications and detection using LISA}.
\newblock {\em Class. Quantum Grav.}, 24:R113--R169, 2007.

\bibitem{davis-etal:71}
M.~Davis, R.~Ruffini, W.H. Press, and R.H. Price.
\newblock {Gravitational radiation from a particle falling radially into
  Schwarzschild black hole}.
\newblock {\em Phys. Rev. Lett.}, 27:1466--1469, 1971.

\bibitem{detweiler:78}
S.L. Detweiler.
\newblock {Black holes and gravitational waves. I. Circular orbits about a
  rotating hole}.
\newblock {\em Astrophys. J.}, 225:687--693, 1978.

\bibitem{regge-wheeler:57}
T.~Regge and J.A. Wheeler.
\newblock {Stability of a Schwarzschild singularity}.
\newblock {\em Phys. Rev.}, 108:1063--1069, 1957.

\bibitem{zerilli:70}
F.J. Zerilli.
\newblock {Gravitational field of a particle falling in a Schwarzschild
  geometry analyzed in tensor harmonics}.
\newblock {\em Phys. Rev. D}, 2:2141--2160, 1970.

\bibitem{teukolsky:73}
S.A. Teukolsky.
\newblock {Perturbations of a rotating black hole. I. Fundamental equations for
  gravitational, electromagnetic, and neutrino-field perturbations}.
\newblock {\em Astrophys. J.}, 185:635--647, 1973.

\bibitem{mino-etal:97}
Y.~Mino, M.~Sasaki, and T.~Tanaka.
\newblock Gravitational radiation reaction to a particle motion.
\newblock {\em Phys. Rev. D}, 55:3457--3476, 1997.

\bibitem{quinn-wald:97}
T.C. Quinn and R.M. Wald.
\newblock Axiomatic approach to electromagnetic and gravitational radiation
  reaction of particles in curved spacetime.
\newblock {\em Phys. Rev. D}, 56:3381--3394, 1997.

\bibitem{quinn:00}
T.C. Quinn.
\newblock Axiomatic approach to radiation reaction of scalar point particles in
  curved spacetime.
\newblock {\em Phys. Rev. D}, 62:064029--1--064029--9, 2000.

\bibitem{gralla-wald:08}
Samuel~E. Gralla and Robert~M. Wald.
\newblock {A Rigorous Derivation of Gravitational Self-force}.
\newblock {\em Class. Quant. Grav.}, 25:205009, 2008.

\bibitem{pound:10a}
A.~Pound.
\newblock Self-consistent gravitational self-force.
\newblock {\em Phys. Rev. D}, 81:024023--1--024023--45, 2010.

\bibitem{barack-ori:00}
L.~Barack and A.~Ori.
\newblock Mode sum regularization approach for the self-force in black hole
  spacetime.
\newblock {\em Phys. Rev. D}, 61:061502--1--061502--5, 2000.

\bibitem{barack:00}
L.~Barack.
\newblock Self-force on a scalar particle in spherically symmetric spacetime
  via mode-sum regularization: Radial trajectories.
\newblock {\em Phys. Rev. D}, 62:084027--1--084027--21, 2000.

\bibitem{barack-burko:00}
L.~Barack and L.~Burko.
\newblock Radiation-reaction force on a particle plunging into a black hole.
\newblock {\em Phys. Rev. D}, 62:084040--1--084040--5, 2000.

\bibitem{barack-lousto:02}
L.~Barack and C.O. Lousto.
\newblock {Computing the gravitational self-force on a compact object plunging
  into a Schwarzschild black hole}.
\newblock {\em Phys. Rev. D}, 66:061502--1--061502--5, 2002.

\bibitem{burko:00c}
L.M. Burko.
\newblock Self-force on a particle in orbit around a black hole.
\newblock {\em Phys. Rev. Lett.}, 84:4529--4532, 2000.

\bibitem{detweiler-etal:03}
S.~Detweiler, E.~Messaritaki, and B.F. Whiting.
\newblock {Self-force of a scalar field for circular orbits about a
  Schwarzschild black hole}.
\newblock {\em Phys. Rev. D}, 67:104016--1--104016--18, 2003.

\bibitem{diazrivera-etal:04}
L.M. Diaz-Rivera, E.~Messaritaki, B.F. Whiting, and S.~Detweiler.
\newblock {Scalar field self-force effects on orbits about a Schwarzschild
  black hole}.
\newblock {\em Phys. Rev. D}, 70:124018--1--124018--14, 2004.

\bibitem{haas-poisson:06}
R.~Haas and E.~Poisson.
\newblock Mode-sum regularization of the scalar self-force: Formulation in
  terms of a tetrad decomposition of the singular field.
\newblock {\em Phys. Rev. D}, 74:004009--1--004009--29, 2006.

\bibitem{barack-sago:07}
L.~Barack and N.~Sago.
\newblock {Gravitational self-force on a particle in circular orbit around a
  Schwarzschild black hole}.
\newblock {\em Phys. Rev. D}, 75:064021--1--064021--25, 2007.

\bibitem{haas:07}
R.~Haas.
\newblock {Scalar self-force on eccentric geodesics in Schwarzschild spacetime:
  a time-domain computation}.
\newblock {\em Phys. Rev. D}, 75:124011--1--124011--17, 2007.

\bibitem{warburton-barack:10}
N.~Warburton and L.~Barack.
\newblock {Self-force on a scalar charge in Kerr spacetime: Circular equatorial
  orbits}.
\newblock {\em Phys. Rev. D}, 81:084039--1--084039--17, 2010.

\bibitem{barack-sago:10}
L.~Barack and N.~Sago.
\newblock Gravitational self-force on a particle in eccentric orbit around a
  schwarzschild black hole.
\newblock {\em Phys. Rev. D}, 81:084021--1--084021--35, 2010.

\bibitem{barack-ori:02}
L.~Barack and A.~Ori.
\newblock {Regularization parameters for the self-force in Schwarzschild
  spacetime: scalar case}.
\newblock {\em Phys. Rev. D}, 66:084022--1--084022--15, 2002.

\bibitem{barack-ori:03a}
L.~Barack and A.~Ori.
\newblock {Regularization parameters for the self-force in Schwarzschild
  spacetime. II. Gravitational and electromagnetic cases}.
\newblock {\em Phys. Rev. D}, 67:024029--1--024029--11, 2003.

\bibitem{barack-ori:03b}
L.~Barack and A.~Ori.
\newblock {Gravitational Self-Force on a Particle Orbiting a Kerr Black Hole}.
\newblock {\em Phys. Rev. Lett.}, 90:111101--1--111101--4, 2003.

\bibitem{detweiler-whiting:03}
S.~Detweiler and B.F. Whiting.
\newblock {Self-force via a Green's function decomposition}.
\newblock {\em Phys. Rev. D}, 67:024025--1--024025--5, 2003.

\bibitem{dirac:38}
P.A.M. Dirac.
\newblock Classical theory of radiating electrons.
\newblock {\em Proc. R. Soc. London, Ser. A}, 167:148, 1938.

\bibitem{barack-golbourn:07}
L.~Barack and D.A. Golbourn.
\newblock Scalar-field perturbations from a particle orbiting a black hole
  using numerical evolution in 2+1 dimensions.
\newblock {\em Phys. Rev. D}, 76:044020--1--044020--20, 2007.

\bibitem{barack-etal:07}
L.~Barack, D.A. Golbourn, and N.~Sago.
\newblock {m-mode regularization scheme for the self-force in Kerr spacetime}.
\newblock {\em Phys. Rev. D}, 76:124036--1--124036--14, 2007.

\bibitem{vega-detweiler:08}
I.~Vega and S.~Detweiler.
\newblock Regularization of fields for self-force problems in curved spacetime:
  Foundations and a time-domain application.
\newblock {\em Phys. Rev. D}, 77:084008--1--084008--14, 2008.

\bibitem{dolan-barack:10}
S.R. Dolan and L.~Barack.
\newblock {Self force via m-mode regularization and 2+1D evolution: Foundations
  and a scalar-field implementation on Schwarzschild}.
\newblock http://arxiv.org/abs/1010.5255, 2010.

\bibitem{vega-etal:09}
I.~Vega, P.~Diener, W.~Tichy, and S.~Detweiler.
\newblock Self-force with (3+1) codes: a primer for numerical relativists.
\newblock {\em Phys. Rev. D}, 80:084021--1--084021--22, 2009.

\bibitem{poissonLLR:04}
E.~Poisson.
\newblock The motion of point particles in curved spacetime.
\newblock {\em Living Rev. Relativity}, 6, 2004.

\bibitem{ottewill-wardell:10}
A.C. Ottewill and B.~Wardell.
\newblock {A Transport Equation Approach to Calculations of Green functions and
  HaMiDeW coefficients}.
\newblock http://arxiv.org/abs/0906.0005, 2010.

\bibitem{ottewill-wardell:09}
A.C. Ottewill and B.~Wardell.
\newblock Quasilocal contribution to the scalar self-force: Nongeodesic motion.
\newblock {\em Phys. Rev. D}, 79:024031--1--024031--10, 2009.

\bibitem{thorne-hartle:85}
K.S. Thorne and J.B. Hartle.
\newblock Laws of motion and precession for black holes and other bodies.
\newblock {\em Phys. Rev. D}, 31:1815--1837, 1985.

\bibitem{zhang:86}
X.-H. Zhang.
\newblock Multipole expansions of the general-relativistic gravitational field
  of the external universe.
\newblock {\em Phys. Rev. D}, 34:991--1004, 1986.

\bibitem{Zenginoglu:2009hd}
Anil Zenginoglu and Manuel Tiglio.
\newblock {Spacelike matching to null infinity}.
\newblock {\em Phys. Rev.}, D80:024044, 2009.

\bibitem{Hod:1999rx}
Shahar Hod.
\newblock {Mode-coupling in rotating gravitational collapse of a scalar field}.
\newblock {\em Phys. Rev.}, D61:024033, 2000.

\bibitem{Pollney:2009ut}
Denis Pollney, Christian Reisswig, Nils Dorband, Erik Schnetter, and Peter
  Diener.
\newblock {The Asymptotic Falloff of Local Waveform Measurements in Numerical
  Relativity}.
\newblock {\em Phys. Rev.}, D80:121502, 2009.

\bibitem{Pollney:2009yz}
Denis Pollney, Christian Reisswig, Erik Schnetter, Nils Dorband, and Peter
  Diener.
\newblock {High accuracy binary black hole simulations with an extended wave
  zone}.
\newblock http://arxiv.org/abs/0910.3803, 2009.

\bibitem{flanagan-hinderer:10}
E.E. Flanagan and T.~Hinderer.
\newblock Transient resonances in the inspirals of point particles into black
  holes.
\newblock http://arxiv.org/abs/1009.0292, 2010.

\end{thebibliography}

\end{document}